\documentclass{aa}

\usepackage[tbtags,fleqn]{amsmath}
\usepackage{amsfonts}
\usepackage{times}
\usepackage{graphics}

\newcommand{\e}{\mathrm e}
\newcommand{\diff}{\mathrm d}
\newcommand{\Lp}{\mathcal L}
\newcommand{\mincir}{\raise
  -2.truept\hbox{\rlap{\hbox{$\sim$}}\raise5.truept \hbox{$<$}\ }}
\newcommand{\magcir}{\raise
  -2.truept\hbox{\rlap{\hbox{$\sim$}}\raise5.truept \hbox{$>$}\ }}

\begin{document}

\title{Smooth maps from clumpy data}
\author{Marco Lombardi and Peter Schneider}
\offprints{M. Lombardi}
\mail{lombardi@astro.uni-bonn.de}
\institute{%
  Instit\"ut f\"ur Astrophysik und Extraterrestrische Forschung,
  Universit\"at Bonn, Auf dem H\"ugel 71, D-53121 Bonn, Germany}
\date{Received ***date***; accepted ***date***}
\abstract{%
  We study an estimator for smoothing irregularly sampled data into a
  smooth map.  The estimator has been widely used in astronomy, owing
  to its low level of noise; it involves a weight function -- or
  smoothing kernel -- $w(\vec\theta)$.  We show that this estimator is
  not unbiased, in the sense that the expectation value of the
  smoothed map is not the underlying process convolved with $w$, but a
  convolution with a modified kernel $w_\mathrm{eff}(\vec\theta)$.  We
  show how to calculate $w_\mathrm{eff}$ for a given kernel $w$ and
  investigate its properties.  In particular, it is found that (1)
  $w_\mathrm{eff}$ is normalized, (2) has a shape `similar' to the
  original kernel $w$, (3) converges to $w$ in the limit of high
  number density of data points, and (4) reduces to a top-hat filter
  in the limit of very small number density of data points.  Hence,
  although the estimator is biased, the bias is well understood
  analytically, and since $w_\mathrm{eff}$ has all the desired
  properties of a smoothing kernel, the estimator is in fact very
  useful.  We present explicit examples for several filter functions
  which are commonly used, and provide a series expression valid in
  the limit of large density of data points.  
  }

\maketitle

%

\section{Introduction}
\label{sec:introduction}

A common problem in astronomy is the smoothing of some irregularly
sampled data into a continuous map.  It is hard to list all possible
cases where such a problem is encountered.  Just to give a few
examples, we mention the determination of the stellar distribution of
our Galaxy, the mapping of column density of dark molecular clouds
from the absorption of background stars, the determination of the
distribution function in globular clusters from radial and, recently,
proper motions of stars, the determination of cosmological large-scale
structures from redshift surveys, and the mass reconstruction in
galaxy clusters from the observed distortion of background galaxies
using weak lensing techniques.  Similarly, one-dimensional data, such
as time series of some events, often need to be smoothed in order to
obtain a real function.

The use of a smooth map is convenient for at least two reasons.
First, a smooth map can be better analyzed than irregularly sampled
data.  Second, if the smoothing is done in a sufficiently coarse way,
the smooth map is significantly less noisy than the individual data.
The drawback related to this last point is a loss in resolution, but
often this is a price that we have to pay in order to obtain results
with a decent signal-to-noise ratio.

In many cases the transformation of irregularly sampled data into a
smooth map follows a standard approach.  A positive weight function,
describing the relative weight of a datum at the position $\vec \theta
+ \vec\phi$ on the point $\vec\theta$, is introduced.  This function
is generally of the form $w(\vec\phi)$, i.e.\ is independent of
absolute position $\vec\theta$ of the point, and actually often
depends only on the separation $| \vec\phi |$.  The weight function
$w(\vec\phi)$ is also chosen so that its value is large when the datum
is close to the point, i.e.\ when $| \vec\phi |$ is small, and
vanishes when $| \vec\phi |$ is large.  Then, the data are averaged
using a weighted mean with the weights given by the function $w$.
More precisely, calling $\hat f_n$ the $n$-th datum obtained at
the position $\vec\theta_n$, the smooth map is defined as
\begin{equation}
  \label{eq:1}
  \tilde f(\vec\theta) = \dfrac{\sum_n \hat f_n w(\vec\theta
  - \vec\theta_n)}{\sum_n w(\vec\theta - \vec\theta_n)} \; .
\end{equation}
It is \textit{reasonable\/} to \textit{assume\/} that this standard
approach works well and produces good results, and the frequent use of
this estimator suggests this is the case.  On the other hand, in our
opinion the properties of the smoothing should be better characterized
by means of rigorous calculations.  Some authors have actually already
studied the smoothing using some approximations, and have obtained
some preliminary results (e.g., Lombardi \& Bertin 1998, van Waerbeke
2000).  To our knowledge, however, the general problem has not been
fully addressed so far and in particular there are no exact results
known.

In this paper we consider in detail the effect of the smoothing on
irregularly sampled data and derive a number of \textit{exact\/}
properties for the resulting map.  We assume that the measurements
$\hat f_n$ are \textit{unbiased\/} estimates of some unknown field
$f(\vec\theta)$ at the positions $\vec\theta_n$, and we study the
expectation value of the map $\tilde f(\vec\theta)$ using an
\textit{ensemble average}, i.e.\ taking the $N$ positions $\bigl\{
\vec\theta_n \bigr\}$ as random variables.  We then show that the
expectation value for the smooth map of Eq.~\eqref{eq:1} is given by
\begin{equation}
  \label{eq:2}
  \bigl\langle \tilde f(\vec\theta) \bigr\rangle = \int f(\vec\theta')
  w_\mathrm{eff}(\vec\theta - \vec\theta') \, \diff^2 \theta' \; .
\end{equation}
Thus, $\bigl\langle \tilde f \bigr\rangle$ is the convolution of the
unknown field $f$ with an \textit{effective weight\/} $w_\mathrm{eff}$
which, in general, differs from the weight function $w$.  We also show
that $w_\mathrm{eff}$ has a ``similar'' shape as $w$ and converges to
$w$ for a large number of objects $N$, but in general $w_\mathrm{eff}$
is broader than $w$.  Moreover, the effective weight is normalized, so
that no signal is ``lost'' or ``created.''\@ Finally, we obtain some
analytical expansions for $w_\mathrm{eff}$, and investigate its
behavior in a number of interesting cases.

A common alternative to Eq.~\eqref{eq:1} is a non-normalized weighted
sum, defined as
\begin{equation}
  \label{eq:3}
  \tilde f(\vec\theta) = \frac{1}{\rho} \sum_{n=1}^N \hat f_n
  w(\vec\theta - \vec\theta_n) \; ,
\end{equation}
where $w$ has been assumed to have unit integral [see below
Eq.~\eqref{eq:6}].  The statistical properties of this estimator can
be easily derived, and in particular we obtain for its expectation
value [cf.\ Eq.~\eqref{eq:3}]
\begin{equation}
  \label{eq:4}
  \bigl\langle \tilde f(\vec\theta) \bigr\rangle = \int f(\vec\theta')
  w(\vec\theta - \vec\theta') \, \diff^2 \theta' \; .
\end{equation}
However, we note that this non-normalized estimator is expected to be
more noisy than the one defined in Eq.~\eqref{eq:1} because of
sampling noise.  For example, even in the case of a flat field
$f(\vec\theta) = 1$ measured without errors (so that $\hat f_n = 1$)
we expect to have a noisy map if we use Eq.~\eqref{eq:3}.  For this
reason, whenever possible the estimator \eqref{eq:1} should be used
instead.

The paper is organized as follows.  In Sect.~\ref{sec:definitions} we
derive some preliminary expressions for the mean value of the map of
Eq.~\eqref{eq:1}.  These results are generalized to a variable number
$N$ of objects in Sect.~\ref{sec:continuous-limit}.  If the weight
function $w(\vec\theta)$ is allowed to vanish, then some peculiarities
arises.  This case is considered in detail in
Sect.~\ref{sec:vanishing-weights}.  Section~\ref{sec:properties-cw} is
dedicated to general properties for the average of $\tilde f$ and
related functions.  In Sect.~\ref{sec:moments-expansion} we take an
alternative approach which can be used to obtain an analytic expansion
for $\tilde f$.  In Sect.~\ref{sec:examples}, we consider three
specific examples of weight functions often used in practice.
Finally, a summary of the results obtained in this paper is given in
Sect.~\ref{sec:conclusions}.  A variation of the smoothing technique
considered in this paper is briefly discussed in
Appendix~\ref{sec:object-intr-weight}.

\section{Definitions and first results}
\label{sec:definitions}

Suppose one wants to measure an unknown field $f(\vec\theta)$, a
function of the ``position'' $\vec\theta$. [What $\vec\theta$ really
means is totally irrelevant for our discussion.  For example,
$\vec\theta$ could represent the position of an object on the sky, the
time of some observation, or the wavelength of a spectral feature.  In
the following, to focus on a specific case, we will assume that
$\vec\theta$ represents a position on the sky and thus we will
consider it as two-dimensional variable.]\@ Suppose also that we can
obtain a total of $N$ \textit{unbiased\/} estimates $\hat f_n$ for $f$
at some points $\bigl\{ \vec\theta_n \bigr\}$, and that each point can
freely span a field $\Omega$ of surface $A$ ($\Omega$ represents the
area of the survey, i.e.\ the area where data are available).  The
points $\bigl\{ \vec\theta_n \bigr\}$, in other words, are taken to be
\textit{independent\/} random variables with uniform probability
distribution and density $\rho = N / A$ inside the set $\Omega$ of
their possible values.  We can then define the smooth map of
Eq.~\eqref{eq:1}, or more explicitly
\begin{equation}
  \label{eq:5}
  \tilde f(\vec\theta) = \dfrac{\sum_{n=1}^N \hat f_n
  w(\vec\theta - \vec\theta_n)}{\sum_{n=1}^N w(\vec\theta -
  \vec\theta_n)} \; .
\end{equation}
In the rest of this paper we study the expectation value $\bigl\langle
\tilde f(\vec\theta) \bigr\rangle$ of $\tilde f(\vec\theta)$ (an
alternative weighting scheme is briefly discussed in
Appendix~\ref{sec:object-intr-weight}).  To simplify the notation we
will assume, without loss of generality, that the weight function
$w(\vec\theta)$ is normalized, i.e.
\begin{equation}
  \label{eq:6}
  \int_\Omega w(\vec\theta) \, \diff^2 \theta = 1 \; .
\end{equation}

In order to obtain the ensemble average of $\tilde f$ we need to
average over all possible measurements at each point, i.e.\ $\hat
f_n$, and over all possible positions $\bigl\{ \vec\theta_n \bigr\}$
for the $N$ points.  The first average is trivial, since $\tilde
f(\vec\theta)$ is linear on the data $\hat f_n$ and the data are
unbiased, so that $\bigl\langle \hat f_n \bigr\rangle =
f(\vec\theta_n)$.  We then have
\begin{align}
  \label{eq:7}
  \bigl\langle \tilde f(\vec\theta) \bigr\rangle = {} & \frac{1}{A^N}
  \int_\Omega \diff^2 \theta_1 \int_\Omega \diff^2 \theta_2
  \dotsi \nonumber\\
  & {} \times \int_\Omega \diff^2 \theta_N \dfrac{\sum_{n=1}^N f(
  \vec\theta_n ) w( \vec\theta -
  \vec\theta_n)}{\sum_{n=1}^N w( \vec\theta -
  \vec\theta_n )} \; .
\end{align}
Relabeling the integration variables we can rewrite this expression as
\begin{align}
  \label{eq:8}
  \bigl\langle \tilde f(\vec\theta) \bigr\rangle = {} & \frac{N}{A^N}
  \int_\Omega \diff^2 \theta_1 \int_\Omega \diff^2 \theta_2
  \dotsi \nonumber\\
  & {} \times \int_\Omega \diff^2 \theta_N \dfrac{f( \vec\theta_1
  ) w( \vec\theta - \vec\theta_1)}{\sum_{n=1}^N
  w( \vec\theta - \vec\theta_n )} \; .
\end{align}
We now define a new random variable
\begin{equation}
  \label{eq:9}
  y(\vec\theta) = \sum_{n=2}^N w( \vec\theta - \vec\theta_n ) \; .
\end{equation}
Note that the sum runs from $n=2$ to $n=N$.  Let us call $p_y(y)$ the
probability distribution for $y(\vec\theta)$.  If we suppose that
$\vec\theta$ is not close to the boundary of $\Omega$, so that the
support of $w(\vec\theta - \vec\theta')$ (i.e.\ the set of points
$\vec\theta'$ where $w(\vec\theta - \vec\theta') \neq 0$) is inside
$\Omega$, then the probability distribution for $y(\vec\theta)$ does
not depend on $\vec\theta$.  We anticipate here that below we will
take the limit of large surveys, so that $\Omega$ tends to the whole
plane, and $A \rightarrow \infty$, $N \rightarrow \infty$, such that
$\rho = N/A$ remains constant.  Since, by definition, the weight
function is assumed to be non-negative, $p_y(y)$ vanishes for $y < 0$.
Analogously, we call $p_w(w)$ the probability distribution for the
weight $w$.  These two probability distributions can be calculated
from the equations
\begin{align}
  \label{eq:10}
  p_w(w) &{} = \frac{1}{A} \int_\Omega \delta\bigl( w - w(\vec\theta)
  \bigr) \, \diff^2 \theta \; , \\
  \label{eq:11}
  p_y(y) &{} = \frac{1}{A^{N-1}} \int_\Omega \diff^2 \theta_2 \dotsi
  \int_\Omega \diff^2 \theta_N \delta( y - w_2 - \dots - w_N )
  \nonumber\\
  &{} = \int_0^\infty \diff w_2 p_w( w_2 ) \int_0^\infty \diff w_3
  p_w( w_3 ) \: \dotsi \nonumber\\
  &\phantom{{} =} {} \times \int_0^\infty \diff w_N p_w( w_N ) \delta(
  y - w_2 - \dots - w_N ) \; ,
\end{align}
where $\delta$ is Dirac's distribution and where we have called $w_n =
w(\vec\theta_n)$.  Note that Eqs.~\eqref{eq:10} and \eqref{eq:11} hold
only if the $N$ points $\bigl\{ \vec\theta_n \bigr\}$ are uniformly
distributed on the area $A$ with density $\rho$, and if there is no
correlation (so that the probability distribution for each point is
$p_{\vec\theta}(\vec\theta_n) = 1/A$).  Moreover, we are assuming here
that the probability distribution for $y(\vec\theta)$ does not depend
on $\vec\theta$.  This is true only if a given configuration of points
$\bigl\{ \vec\theta_n \bigr\}$ has the same probability as the
translated set $\bigl\{ \vec\theta_n + \vec\phi \bigr\}$.  This
translation invariance, clearly, cannot hold exactly for finite fields
$\Omega$; on the other hand, again, as long as $\vec\theta$ is far
from the boundary of the field, the probability distribution for
$y(\vec\theta)$ is basically independent of $\vec\theta$.  Note that
in case of a field with masks, we also have to exclude in our analysis
points close to the masks.

Using $p_y$ we can rewrite Eq.~\eqref{eq:8} in a more compact form:
\begin{align}
  \label{eq:12}
  \bigl\langle \tilde f(\vec\theta) \bigr\rangle = {} & \rho
  \int_\Omega \diff^2 \theta_1 f( \vec\theta_1 )
  w( \vec\theta - \vec\theta_1) \nonumber\\
  & {} \times \int_0^\infty \frac{p_y(y)}{w(\vec\theta - \vec\theta_1)
  + y} \, \diff y \; ,
\end{align}
where, we recall, $\rho = N/A$ is the density of objects.  For the
following calculations, it is useful to write this equation as
\begin{equation}
  \label{eq:13}
  \bigl\langle \tilde f(\vec\theta) \bigr\rangle = 
  \int_\Omega \diff^2 \theta' f(\vec\theta') w(\vec\theta -
  \vec\theta') C\bigl( w(\vec\theta - \vec\theta') \bigr) \, \diff^2
  \theta' \; ,
\end{equation}
where $C(w)$ the \textit{correcting factor}, defined as
\begin{equation}
  \label{eq:14}
  C(w) = \rho \int_0^\infty \frac{p_y(y)}{w + y} \, \diff y \; .
\end{equation}
Finally, we will often call the combination
$w_\mathrm{eff}(\vec\theta) = w(\vec\theta) C\bigl( w(\vec\theta)
\bigr)$, which enters Eq.~\eqref{eq:13}, \textit{effective weight}.  

Interestingly, Eq.~\eqref{eq:13} shows that the relationship between
$\bigl\langle \tilde f(\vec\theta) \bigr\rangle$ and $f(\vec\theta)$
is a simple convolution with the kernel $w_\mathrm{eff}$.  From the
definition \eqref{eq:5}, we can also see that this kernel is
normalized, in the sense that
\begin{equation}
  \label{eq:15}
  \int_\Omega \bigl\langle \tilde f(\vec\theta) \bigr\rangle \, \diff^2
  \theta = \int_\Omega f(\vec\theta) \, \diff^2 \theta \; .
\end{equation}
In fact, if we consider a ``flat'' signal, for instance $f(\vec\theta)
= 1$, we clearly obtain $\bigl\langle \tilde f(\vec\theta)
\bigr\rangle = 1$.  On the other hand, from the properties of
convolutions, we know that the ratio between the l.h.s.\ and the
r.h.s.\ of Eq.~\eqref{eq:15} is constant, independent of the function
$f(\vec\theta)$.  We thus deduce that this ratio is $1$, i.e.\ that
Eq.~\eqref{eq:15} holds in general.  The normalization of
$w_\mathrm{eff}$ will be also proved below in
Sect.~\ref{sec:normalization} using analytical techniques.

If $p_y(y)$ is available, Eq.~\eqref{eq:12} can be used to obtain the
expectation value for the smoothed map $\tilde f$.  In order to obtain
an expression for $p_y$ we use Markov's method (see, e.g.,
Chandrasekhar 1943; see also Deguchi \& Watson 1987 for an application
to microlensing studies).  Let us define the Laplace transforms of
$p_y$ and $p_w$:
\begin{align}
  \label{eq:16}
  W(s) & {} = \Lp[p_w](s) = \int_0^\infty \e^{-sw} p_w(w) \, \diff w
  \nonumber\\
  & {} = \frac{1}{A} \int_\Omega \e^{-sw(\vec\theta)} \, \diff^2
  \theta \; , \\
  \label{eq:17}
  Y(s) & {} = \Lp[p_y](s) = \int_0^\infty \e^{-sy} p_y(y) \, \diff y
  = \bigl[ W(s) \bigr]^{N-1} \; .
\end{align}
Hence $p_y$ can in principle be obtained from the following scheme.
First, we evaluate $W(s)$ using Eq.~\eqref{eq:16}, then we calculate
$Y(s)$ from Eq.~\eqref{eq:17}, and finally we back-transform this
function to obtain $p_y(y)$.

\section{Continuous limit}
\label{sec:continuous-limit}

So far we have considered a finite set $\Omega$ and a fixed number of
objects $N$.  In reality, one often deals with a non-constant number
of objects, so that $N$ is itself a random variable.  [For example, if
the objects we are studying are galaxies and if $\Omega$ is a small
field, the expected number of galaxies in our field will follow a
Poissonian distribution with mean value $\rho A$, where $\rho$ is the
density of detectable galaxies in the sky.]\@ Clearly, when we observe
a field $\Omega$ we will obtain a particular value for the number of
objects $N$ inside the field.  However, in order to obtain more
general results, it is convenient to consider an ensemble average, and
take the number of observed objects as a random variable; the results
obtained, thus, will be averaged over all possible values of $N$.

A way to include the effect of a variable $N$ in our framework is to
note that, although we are observing a small area on the sky, each
object could in principle be located at any position of the whole sky.
Hence, instead of taking $N$ as a random variable and $A$ fixed, we
consider larger and larger areas of the sky and take the limit $A
\rightarrow \infty$.  In doing this, we keep the object density $\rho
= N / A$ constant.  It is easily verified that the two methods (namely
$A$ fixed and $N$ Poissonian random variable with mean $\rho A$, or
rather $A \rightarrow \infty$ with $\rho = A / N$ fixed), lead to the
same results.  In the following, however, we will take the latter
scheme, and let $A$ goes to infinity; correspondingly we take $\Omega$
as the whole plane.

From Eq.~\eqref{eq:16} we see that $W(s)$ is proportional to $1/A$.
For this reason it is convenient to define the function
\begin{equation}
  \label{eq:18}
  Q(s) = \int_\Omega \left[ \e^{-s w(\vec\theta)} - 1 \right] \,
  \diff^2 \theta \; .
\end{equation}
If $A$ is finite, we have $Q(s) = A W(s) - A$.  Thus we can write, in
the limit $A \rightarrow \infty$,
\begin{equation}
  \label{eq:19}
  Y(s) = \lim_{N \rightarrow \infty} \left[ 1 + \frac{Q(s) \rho}{N}
  \right]^{N-1} = \e^{\rho Q(s)} \; .
\end{equation}
This equation replaces Eq.~\eqref{eq:17} when $N \rightarrow \infty$.

In order to further simplify the expression for the correcting factor
$C(w)$, we rewrite its definition as
\begin{equation}
  \label{eq:20}
  C(w) = \rho \int_0^\infty \frac{\zeta_w(x)}{x} \, \diff x \; ,
\end{equation}
where $x = y + w$, and the function $\zeta_w$ is defined as
\begin{equation}
  \label{eq:21}
  \zeta_w(x) = \mathrm{H}(x - w) p_y(x - w) \; .
\end{equation}
Here $\mathrm{H}(x - w)$ is the Heaviside function at the position
$w$, i.e.
\begin{equation}
  \label{eq:22}
  \mathrm{H}(x) = 
  \begin{cases}
    0 & \text{if $x < 0 \; ,$} \\
    1 & \text{otherwise.}
  \end{cases}
\end{equation}
The Laplace transform of $\zeta_w$ can be written as
\begin{equation}
  \label{eq:23}
  Z_w(s) = \Lp[\zeta_w](s) = \e^{-ws} Y(s) \; .
\end{equation}
In reality, for $C(w)$ we need to evaluate an integral over $\zeta_w(x)
/ x$.  From the properties of the Laplace transform we have
\begin{equation}
  \label{eq:24}
  \Lp\bigl[ \zeta_w(x) / x \bigr](s) = \int_s^\infty Z_w(s') \, \diff s'
  \; ,
\end{equation}
and thus we find
\begin{align}
  \label{eq:25}
  C(w) & {} = \rho \Lp\bigl[ \zeta_w(x) / x \bigr](0) = \rho
  \int_0^\infty Z_w(s') \, \diff s' \nonumber\\
  & {} = \rho \int_0^\infty \e^{-w s'} Y(s') \, \diff s' = \Lp[\rho
  Y](w) \; .
\end{align}
This important result, together with Eqs.~\eqref{eq:18} and
\eqref{eq:19}, can be used to readily evaluate the correcting factor.
We note that, for our purposes, there is no need to evaluate the
probability distribution $p_y$ any more.  This prevents us from
calculating any inverse Laplace transform.

\section{Vanishing weights}
\label{sec:vanishing-weights}

Since $w(\vec\theta) \ge 0$ for every $\vec\theta \in \Omega$, $y$ is
non-negative, so that $p_y(y) = 0$ if $y < 0$.  In principle, however,
we cannot exclude that the case $y = 0$ has a finite probability.  In
particular, for finite support weight functions, $p_y(y)$ could
include the contribution from a Dirac delta distribution centered on
zero.

Since $y$ is the sum of weights at different positions, $y = 0$ can
have a finite probability only if $w(\vec\theta)$ vanishes at some
points.  In turn, if $P_0 = P(y = 0)$ is finite we could encounter
situations where the numerator and the denominator of Eq.~\eqref{eq:5}
vanish.  In such cases we could not even define our estimator $\tilde
f$.  We also note that, in the continuous limit, $P_0$ is also the
probability of having vanishing denominator in Eq.~\eqref{eq:5}.  As a
result, if $P_0 > 0$, we have a finite probability of being unable to
evaluate our estimator!

So far we have implicitly assumed that $P_0$ vanishes.  Actually, we
can explicitly evaluate this probability using the expression
\begin{equation}
  \label{eq:26}
  P(y=0) = \lim_{y \rightarrow 0} \int_0^y p_y(y') \, \diff y' \; .
\end{equation}
From the properties of the Laplace transform this expression can be
written as
\begin{equation}
  \label{eq:27}
  P(y=0) = \lim_{s \rightarrow \infty} Y(s) = \lim_{s \rightarrow
  \infty} \e^{\rho Q(s)} \; .
\end{equation}
We now observe that
\begin{equation}
  \label{eq:28}
  \lim_{s \rightarrow \infty} \left[ \e^{-s w(x)} - 1 \right] = 
  \begin{cases}
    -1 & \text{if $w \neq 0 \; ,$} \\
    0  & \text{otherwise.}
  \end{cases}
\end{equation}
As a result, in the limit considered, $-Q(s)$ approaches the area of
the support of $w$. Calling this area $\pi_w$, we find finally
\begin{equation}
  \label{eq:29}
  P(y=0) = \e^{-\rho \pi_w} \; .
\end{equation}
In the case where $w(\vec\theta)$ has infinite support (for example if
$w$ is a Gaussian), $P_0$ vanishes as expected.

The result expressed by Eq.~\eqref{eq:29} can actually be derived
using a more direct approach.  Since $w(\vec\theta) \geq 0$, the
condition $y = 0$ requires that all weights except $w(
\vec\theta_1 )$ are vanishing.  In turn, this happens only if
there is no object inside a neighborhood of the point considered.
Since the area of this neighborhood is $\pi_w$, the number of objects
inside this set follows a Poisson distribution of mean $\rho \pi_w$,
and thus the probability of finding no object is given precisely by
Eq.~\eqref{eq:29}.

As mentioned before, $P_0 > 0$ is a warning that is same cases we
cannot evaluate our estimator $\tilde f$.  In order to proceed in our
analysis allowing for finite-support weight functions, we decide to
\textit{explicitly exclude in our calculations cases where $w + y =
  0$\/}: In such cases, in fact, we could not define $\tilde f$.  In
practice when smoothing the data we would mark as ``bad points'' the
locations $\vec\theta$ where $\tilde f(\vec\theta)$ is not defined.
In taking the ensemble average, then, we would exclude, for each
possible configuration $\bigl\{ \vec\theta_n \bigr\}$, the bad points.
In order to apply this prescription we need to modify $p_y$, the
probability distribution for $y$, and explicitly exclude cases with
$w + y = 0$.  In other words, we define a new probability distribution
for $y$ given by
\begin{equation}
  \label{eq:30}
  \tilde p_y(y) = 
  \begin{cases}
    p_y(y) & \text{if $w \neq 0 \; ,$} \\
    \bigl[ p_y(y) - P_0 \delta(y) \bigr] / (1 - P_0) & \text{if $w = 0
    \; .$}
  \end{cases}
\end{equation}
Hence, if $w = 0$, we set to zero $P(y = 0)$ and then renormalize the
distribution.  An important consequence of the new prescription is
that \textit{the probability distribution for $y$ no longer is
  independent of $w$}.  In fact, the probability $P(y = 0)$ vanishes
for $w = 0$, while is finite (for a finite-support weight) if $w \neq
0$.  Using Eq.~\eqref{eq:30} in the definition of $Y(s)$ we then find
(see Eqs.~\eqref{eq:17} and \eqref{eq:19})
\begin{equation}
  \label{eq:31}
  Y(s) = \e^{\rho Q(s)} - \Delta(w) \frac{P_0}{1 - P_0} \left[
  1 - \e^{\rho Q(s)} \right] \; ,
\end{equation}
where $\Delta(w)$ is $1$ for $w = 0$ and vanishes for $w > 0$.
Equation~\eqref{eq:31} replaces Eq.~\eqref{eq:19} for the cases when
$P_0 > 0$, i.e.\ for weight functions with finite support.

In order to implement the new requirement $w + y \neq 0$, we still
need to modify Eq.~\eqref{eq:12} [or, equivalently,
Eq.~\eqref{eq:14}].  In fact, in deriving that result, we have assumed
that all objects can populate the area $\Omega$ with uniform
probability and in fact we have used a factor $1/A^N$ to normalize
Eq.~\eqref{eq:7}.  Now, however, we must take into account the fact
that objects cannot make a ``void'' around the point $\vec\theta$.  As
a result, we need a further factor $1/(1-P_0)$ in front of
Eqs.~\eqref{eq:12} and \eqref{eq:14}.  In summary, the new set of
equations is
\begin{align}
  \label{eq:32}
  Q(s) &{} = \int_\Omega \left[ \e^{-s w(\vec\theta)} - 1 \right] \,
  \diff^2 \theta \; , \\
  \label{eq:33}
  Y(s) &{} = \e^{\rho Q(s)} - \Delta(w) \frac{P_0}{1 - P_0} \left[
    1 - \e^{\rho Q(s)} \right] \; , \\
  \label{eq:34}
  C(w) &{} = \frac{\rho}{1 - P_0} \int_0^\infty \e^{-w s} Y(s) \,
    \diff s \; ,
\end{align}
where, we recall, $P_0 = \e^{-\rho \pi_w}$ vanishes for weight
functions with infinite support $\pi_w$.

\section{Properties}
\label{sec:properties-cw}

In this section we show a number of interesting properties of $C(w)$
and $\bigl\langle \tilde f(\vec\theta) \bigr\rangle$.

\subsection{Normalization}
\label{sec:normalization}

As already shown [see above Eq.~\eqref{eq:13}], $\bigl\langle \tilde
f(\vec\theta) \bigr\rangle$ is a simple convolution of $f(\vec\theta)$
with $w_\mathrm{eff}$.  The normalization $I$ of this convolution can
be obtained from the expression
\begin{equation}
  \label{eq:35}
  I = \int_\Omega w_\mathrm{eff}(\vec\theta) \, \diff^2 \theta =
  \int_\Omega w(\vec\theta) C\bigl( w(\vec\theta) \bigr) \, \diff^2
  \theta \; .
\end{equation}
From Eq.~\eqref{eq:34} we find
\begin{equation}
  \label{eq:36}
  I = \frac{\rho}{1 - P_0} \int_\Omega \diff^2 \theta \int_0^\infty
  w(\vec\theta) \e^{-s w(\vec\theta)} Y(s) \, \diff s \; .
\end{equation}
We first note that, in the general case, $Y(s)$ \textit{depends\/} on
$w(\vec\theta)$.  On the other hand, since the second term of $Y(s)$
is proportional to $\Delta(w)$ [see Eq.~\eqref{eq:33}], it does not
contribute to the integral.  We can then evaluate first the integral
over $\vec\theta$, obtaining
\begin{equation}
  \label{eq:37}
  \int_\Omega w(\vec\theta) \e^{-s w(\vec\theta)} \, \diff^2 \theta =
  - Q'(s) \; ,
\end{equation}
and thus we have
\begin{equation}
  \label{eq:38}
  I = - \frac{\rho}{1 - P_0} \int_0^\infty Q'(s) \e^{\rho Q(s)} \,
  \diff s = \left. \frac{1}{1 - P_0} \e^{\rho Q(s)} \right|_0^\infty = 1 \; .
\end{equation}
Hence we conclude that the estimator \eqref{eq:5} is correctly
normalized.

\subsection{Scaling}
\label{sec:scaling}

It is easily verified that a simple scaling property holds for the
effective weight.  Suppose that we evaluate
$w_\mathrm{eff}(\vec\theta)$ for a given weight function
$w(\vec\theta)$ and density $\rho$.  If we rescale the weight function
into $k^2 w(k \vec\theta)$, and the density into $k^2 \rho$, where $k$
is a positive factor, the corresponding effective weight is also
rescaled similarly to $w$, i.e.\ $k^2 w_\mathrm{eff}( k \vec\theta)$.

We also recall here that, although a normalized weight function has
been assumed [see Eq.~\eqref{eq:6}], all results are clearly
independent of the normalization of $w$.  Hence, we can also consider
a trivial scaling property: The weight function $k w(\vec\theta)$ has
effective weight $w_\mathrm{eff}(\vec\theta)$ independent of $k$.

We anticipate that the effective weight is very close to the original
weight $w$ for large densities $\rho$ (see below
Sect.~\ref{sec:limit-large-dens}).  The scaling properties discussed
above suggest that the shape of the effective weight is actually
controlled by the expected number of objects for which the weight is
significantly different from zero.  This justifies the definition of
the \textit{weight area\/} $\mathcal{A}$ of $w$:
\begin{equation}
  \label{eq:39}
  \mathcal{A} = \biggl[ \int_\Omega w(\vec\theta) \,
  \diff^2 \theta \biggr]^2 \biggm/ \biggl[ \int_\Omega \bigl[
  w(\vec\theta) \bigr]^2 \, \diff^2 \theta \biggr] \; .
\end{equation}
The first factor in this definition ensures that $\mathcal{A}$ does
not depend on the normalization of $w(\vec\theta)$.  It is easily
verified that $\mathcal{A} = \pi_w$ for a top-hat weight function.
Correspondingly, we define the \textit{weight number of objects\/} as
$\mathcal{N} = \rho \mathcal{A}$.  Clearly, this quantity is left
unchanged by the scalings considered above.  Similar definitions can
be provided for the effective weight $w_\mathrm{eff}$.  Explicitely,
we have
\begin{equation}
  \label{eq:40}
  \mathcal{A}_\mathrm{eff} = \biggl[ \int_\Omega
  w_\mathrm{eff}(\vec\theta) \, \diff^2 \theta \biggr]^2 \biggm/
  \biggl[ \int_\Omega \bigl[ w_\mathrm{eff}(\vec\theta) \bigr]^2 \,
  \diff^2 \theta \biggr] \; ,
\end{equation}
and $\mathcal{N}_\mathrm{eff} = \rho \mathcal{A}_\mathrm{eff}$.

Numerical calculations for $w_\mathrm{eff}$ show clearly that
$\mathcal{N}$, rather than $\rho$, is the key factor that controls
how close the effective weight $w_\mathrm{eff}(\vec\theta)$ is to
$w(\vec\theta)$ (cf.\ Figs.~\ref{fig:1}, \ref{fig:3}, and
\ref{fig:4}).

\subsection{Behavior of $w C(w)$}
\label{sec:behavior-w-cw}

We can easily study the general behavior of $w_\mathrm{eff}$.  We
first consider the case of infinite-support weights ($P_0 = 0$).

We first note that $Y(s) > 0$ for every $s$, and thus $C(w)$ decreases
as $w$ increases.  On the other hand, we have
\begin{equation}
  \label{eq:41}
  w_\mathrm{eff} = w C(w) = \rho Y(0) + \rho \Lp[Y'](w) \; .
\end{equation}
Since $Y'(s) = \rho Q'(s) Y(s) < 0$ is negative, we have that $w C(w)$
increases with $w$.  This result shows that the effective weight
$w_\mathrm{eff}$ follows the general shape of the weight $w$, as
expected.  Moreover, the fact that $C(w)$ is monotonically decreasing
implies that the effective weight $w_\mathrm{eff}$ is ``broader'' than
$w$.  For instance, we can say that there is a value $w_1$ for the
weight such that $w_\mathrm{eff} = w C(w)$ is not larger than $w$ for
$w > w_1$, and not smaller than $w$ for $w < w_1$.  In fact, since
$C(w)$ is monotonic, the equation $C(w) = 1$ can have at most one
solution.  On the other hand, if $C(w) < 1$ (respectively, if $C(w) >
1$) for all $w$, then $w_\mathrm{eff}(\vec\theta) < w(\vec\theta)$
($w_\mathrm{eff}(\vec\theta) > w(\vec\theta)$) for all $\vec\theta$.
This inequality, however, cannot be true since both $w$ and
$w_\mathrm{eff}$ are normalized.

\begin{figure}[!t]
  \begin{center}
    \resizebox{\hsize}{!}{\includegraphics{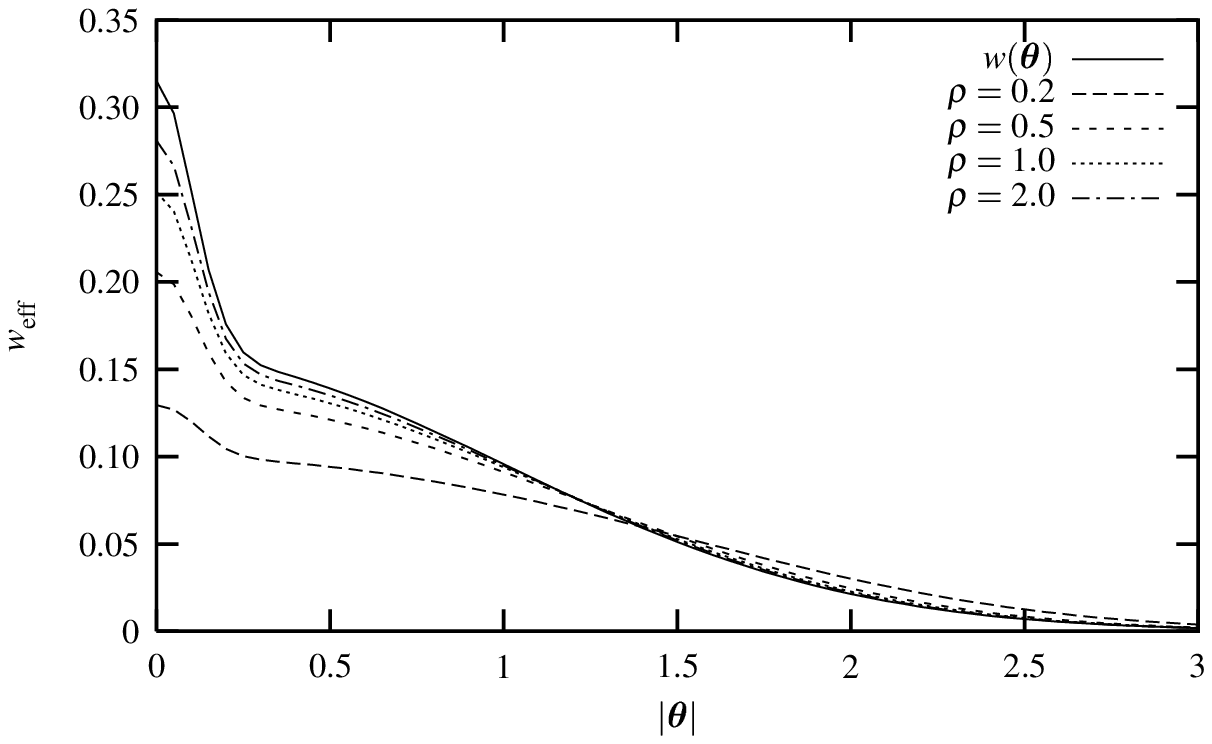}}
    \caption{The effective weight $w_\mathrm{eff}$ never exceed
      $\rho$.  The original weight $w(\vec\theta)$ is the combination
      of two Gaussians with different widths ($\sigma_1 = 1$ and
      $\sigma_2 = 0.1$).  The central peak is severely depressed for
      relative low-densities.  Note that the weight area is
      $\mathcal{A} \simeq 12.2$, so that a density of $\rho = 0.2$
      corresponds to $\mathcal{N} \simeq 2.4$.}
    \label{fig:1}
  \end{center}
\end{figure}

The property just shown can be used to derive a relation between the
weight area $\mathcal{A}$ and the effective weight area
$\mathcal{A}_\mathrm{eff}$.  Let us evaluate the integral
\begin{equation}
  \label{eq:42}
  D = \int_\Omega \bigl[ w(\vec\theta) + w_\mathrm{eff}(\vec\theta) - 2
  w_1 \bigr] \bigl[ w(\vec\theta) - w_\mathrm{eff}(\vec\theta) \bigr]
  \, \diff^2 \theta \; .
\end{equation}
This quantity is positive, since the integrand is positive (this is
easily verified by distinguish the two cases $w(\vec\theta) > w_1$ and
$w(\vec\theta) < w_1$, and by noting that both factors in the
integrand have the same sign).  On the other hand, if we expand the
product we obtain
\begin{align}
  \label{eq:43}
  0 < D & {} = \int_\Omega \bigl[ w(\vec\theta) \bigr]^2 \, \diff^2
  \theta - \int_\Omega \bigl[ w_\mathrm{eff}(\vec\theta) \bigr]^2 \,
  \diff^2 \theta \notag\\
  & \phantom{{} = {}} {} - 2 w_1 \int_\Omega \bigl[ w(\vec\theta) -
  w_\mathrm{eff}(\vec\theta) \bigr] \, \diff^2 \theta \notag\\
  & {} = \mathcal{A}^{-1} - \mathcal{A}_\mathrm{eff}^{-1} \; .
\end{align}
The last relation holds because of the normalization of
$w(\vec\theta)$ and $w_\mathrm{eff}(\vec\theta)$.  We thus have shown
that the effective weight area $\mathcal{A}_\mathrm{eff}$ is always
larger than the original weight area $\mathcal{A}$; analogously we
have $\mathcal{N}_\mathrm{eff} > \mathcal{N}$.  This, clearly, is
another indication that $w_\mathrm{eff}$ is ``broader'' than $w$.

It is also interesting to evaluate the limits of $w C(w)$ for small
and large values of $w$.  We have
\begin{equation}
  \label{eq:44}
  \lim_{w \rightarrow \infty} w C(w) = \rho Y(0) = \rho \; .
\end{equation}
Since we know that $w C(w)$ is monotonic, $\rho$ is also a superior
limit for the effective weight function.  In other words, even if $w$
has high peaks, the effective weight $w_\mathrm{eff}$ (that, we
recall, is normalized) will never exceed the value $\rho$ (see
Fig.~\ref{fig:1}).  We stress here that, since $w_\mathrm{eff}$ is
normalized, its maximum value [which, in virtue of Eq.~\eqref{eq:44}
does not exceed $\rho$] is a significant parameter.  For example,
using the relation $w_\mathrm{eff}(\vec\theta) < \rho$ in the
definition of $\mathcal{A}_\mathrm{eff}$ we find immediately
\begin{equation}
  \label{eq:45}
  \mathcal{A}_\mathrm{eff}^{-1} = \int_\Omega \bigl[
  w_\mathrm{eff}(\vec\theta) \bigr]^2 \, \diff^2 \theta < \rho
  \int_\Omega w_\mathrm{eff}(\vec\theta) \, \diff^2 \theta = \rho \; ,
\end{equation}
or, equivalently, $\mathcal{N}_\mathrm{eff} > 1$.  In other words, no
matter how small $\mathcal{A}$ is, the effective weight will always
``force'' to use at least one object.

Equation~\eqref{eq:44} suggests also a \textit{local\/}
order-of-magnitude check for the effective weight: Assuming
$w(\vec\theta)$ normalized, we expect the effective weight to be
significantly different from $w$ for points where $w(\vec\theta)$ is
of the order of $\rho$ or larger.  Equation~\eqref{eq:45}, instead,
provides a \textit{global\/} criterion: The effective weight will be
significantly broader than $w$ if $\mathcal{N}$ is of the order of
unity or smaller.  As anticipated above, thus, $\mathcal{N}$ is the
real key factor that controls the shape of $w_\mathrm{eff}$ with
respect to $w$.

The other limit for the effective weight is
\begin{equation}
  \label{eq:46}
  \lim_{w \rightarrow 0^+} w C(w) = \rho \lim_{s \rightarrow
  \infty} Y(s) = 0 \; .
\end{equation}
Thus, as expected, the effective weight $w_\mathrm{eff}$ vanishes as
$w$ vanishes.

If $w$ has finite support, then the situation is slightly different.
Given the definition of $Y(s)$, we expect for $C(w)$ a discontinuity
for $w = 0$.  Apart from this difference, the behavior of $C(w)$ and
of $w C(w)$ is similar to the case considered above, namely $C(w)$ is
monotonically decreasing and $w C(w)$ is increasing.  We also have
\begin{equation}
  \label{eq:47}
  \lim_{w \rightarrow \infty} w C(w) = \frac{\rho}{1 - P_0} \; ,
\end{equation}
which is similar to Eq.~\eqref{eq:44}.  Note that, as expected from
the normalization of $w_\mathrm{eff}$, $\pi_w \rho / (1 - P_0) \ge 1$
for all densities.  We also find
\begin{equation}
  \label{eq:48}
  \lim_{w \rightarrow 0^{+}} w C(w) = \frac{\rho P_0}{1 - P_0} \; .
\end{equation}
In other words, the effective weight $w C(w)$ \textit{does not
  vanish\/} for small $w$.  If, instead, we take $w = 0$, then we have
$w C(w) = 0$.  This discontinuity, related to the term $\Delta(w)$ in
Eq.~\eqref{eq:33}, is a consequence of a number of properties for the
effective weight: (i) $w_\mathrm{eff}$ is normalized; (ii)
$w_\mathrm{eff}$ is broader than $w$; (iii) $w_\mathrm{eff}$ has the
same support as $w$.  We thus are forced to have a discontinuity for
the effective weight.
  
The result obtained above is also convenient for simplifying equations
for finite-support weight functions.  In fact, for $w > 0$ we can
clearly drop the last term of Eq.~\eqref{eq:33}, thus recovering
Eq.~\eqref{eq:19}; if, instead, $w = 0$, we can directly set $w C(w) =
0$, without further calculations.

\subsection{Limit of high and low densities}
\label{sec:limit-large-dens}

The final result considered in this section is the behavior of the
correcting factor $C(w)$ in the limit $\rho \rightarrow \infty$ and
$\rho \rightarrow 0$.

We observe that $Q(s) \leq 0$ and moreover $\left\lvert Q(s)
\right\rvert$ increases with $s$.  Thus, if $s$ is large, $\e^{\rho
  Q(s)}$ vanishes quickly when $\rho \rightarrow \infty$.  As a
result, in the limit of large densities we are mainly interested in
$Q(s)$ with $s$ small.  Expanding $Q(s)$ around $s = 0^+$, we find
\begin{equation}
  \label{eq:49}
  Q(s) = \sum_{k=1}^\infty \frac{(-1)^k s^k S_k}{k!} \; ,
\end{equation}
where $S_k$ is the $k$-th moment of $w$:
\begin{equation}
  \label{eq:50}
  S_k = \int_\Omega \bigl[ w(\vec\theta) \bigr]^k \, \diff^2 \theta \;
  .
\end{equation}
The normalization of $w$ clearly implies $S_1 = 1$.  Assuming $w > 0$,
to first order we have then $Y(s) \simeq \e^{-s \rho}$, so that
\begin{equation}
  \label{eq:51}
  C(w) \simeq \frac{1}{1 - P_0} \int_0^\infty \rho \e^{-s w} \e^{-s
  \rho} \, \diff s = \frac{1}{1 - P_0} \frac{\rho}{\rho + w}
  \; .
\end{equation}
This expression gives the correcting factor at the first order in
$1/\rho$.  If $w$ has infinite support the expression reduces to $C(w)
= 1 / \bigl[ 1 + w/\rho\bigr]$.  Note that the quantity $w/\rho \sim
1/\mathcal{N}$ is related the \textit{weight number}, i.e.\ the
expected number of objects which contribute significantly to the
signal, for which the weight is not exceedingly small.  Finally, at
the zero order in the limit $\rho \rightarrow \infty$, $C(w)$
converges to unity.  In this case the map $\tilde f(\vec\theta)$ is
expected to be a smoothing of $f(\vec\theta)$ with the same weight
function $w$.

In the limit $\rho \rightarrow 0$, we have $Y(s) \rightarrow 1$, and
thus
\begin{equation}
  \label{eq:52}
  C(w) \simeq \frac{\rho}{1 - P_0} \frac{1}{w} \; .
\end{equation}
In the same limit, $P_0 \simeq 1 - \pi_w \rho$, so that we find
\begin{equation}
  \label{eq:53}
  w_\mathrm{eff} = w C(w) \simeq \frac{1}{\pi_w} \; .
\end{equation}
Recalling that $w C(w)$ vanishes for $w = 0$, we conclude that the
effective weight converges to a top-hat function with support
$\pi_w$ if $w(\vec\theta)$ has finite-support (see Fig.~\ref{fig:4}
for an example).

\section{Moments expansion}
\label{sec:moments-expansion}

\begin{figure}[!t]
  \begin{center}
    \resizebox{\hsize}{!}{\includegraphics{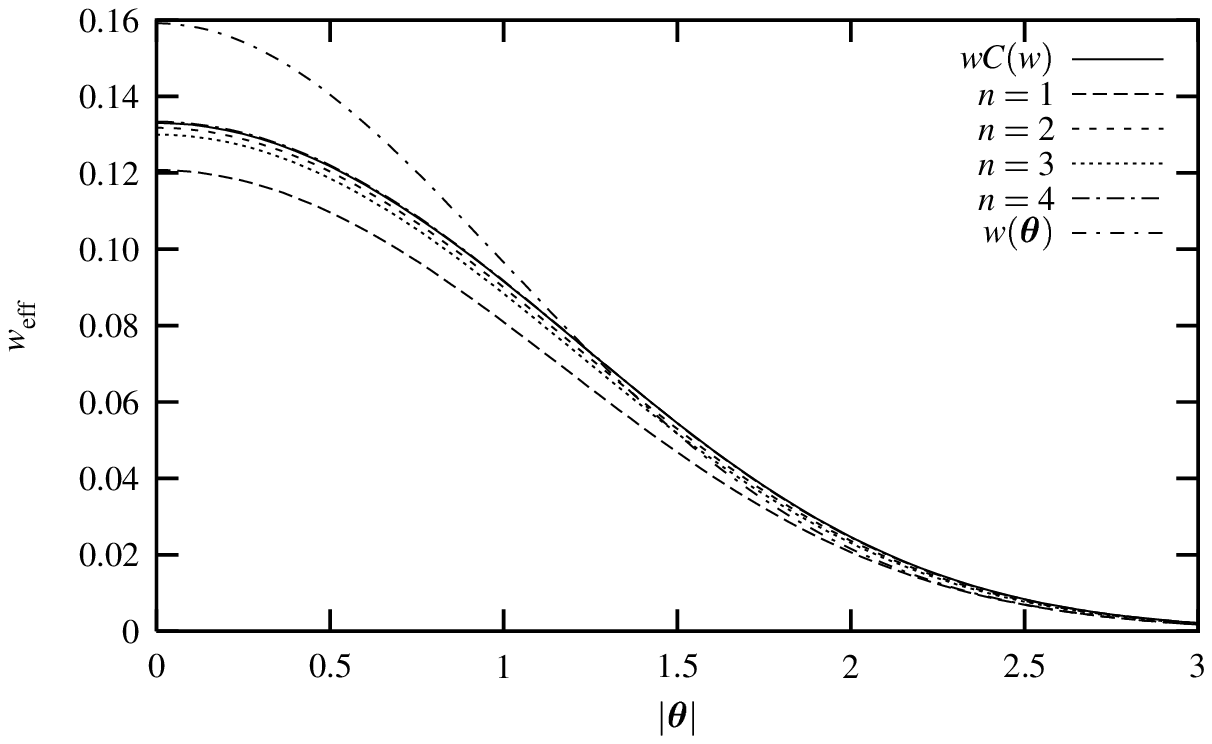}}
    \caption{The effective weight $w_\mathrm{eff}$ can be well
      approximated using the expansion \eqref{eq:66}.  This plot shows
      the behavior of the $n$-th order expansion for a Gaussian weight
      function with unit variance (see Eq.~\eqref{eq:75}).  The
      density used, $\rho = 0.5$, corresponds to $\mathcal{N} \simeq
      6.3$.  The convergence is already extremely good at the second
      order; for larger values of $\mathcal{N}$ the expansion
      converges more rapidly.}
    \label{fig:2}
  \end{center}
\end{figure}

In the last section we have obtained an analytical expansion for
$Q(s)$ which has then been used to obtain a first approximation for
the correction function $C(w)$ valid for large densities $\rho$.
Unfortunately, we have been able to obtain a simple result for $C(w)$
only to first order.  Already at the second order, in fact, the
correcting function would result in a rather complicated expression
involving the error function erf.

Actually, there is a simpler approach to obtain an expansion of $C(w)$
at large $\rho$ using the moments of the random variable $y$.  Given
the definition \eqref{eq:9} for $y$, we expect that $\bar y \equiv
\langle y \rangle$, the average value of this random variable,
increases linearly with the density $\rho$ of objects.  Similarly, for
large $\rho$ the relative scatter $\bigl\langle (y - \bar y)^2
\bigr\rangle / \bar y^2$ is expected to decrease.  In fact, $y$ is the
sum of several independent random variables, and thus, in virtue of
the central limit theorem, it must converge to a Gaussian random
variable with appropriate average and variance.

Since the relative variance of $y$ decreases with $\rho$, we can
expand $y$ in the denominator in Eq.~\eqref{eq:14}, obtaining
\begin{align}
  \label{eq:54}
  C(w) & {} = \rho \int_0^\infty \! \diff y \, \frac{p_y(y)}{\bar y + w}
  \sum_{k=0}^\infty (-1)^k \left( \frac{y - \bar y}{\bar y + w}
  \right)^k \nonumber\\
  & {} = \rho \sum_{k=0}^\infty (-1)^k \frac{1}{(\bar y + w)^{k+1}}
  \bigl\langle (y - \bar y)^k \bigr\rangle \; ,
\end{align}
where we have used the definition of the moments of $y$:
\begin{align}
  \label{eq:55}
  \bar y = \langle y \rangle & {} = \int_0^\infty p_y(y) y \, \diff y \; ,
  \\
  \label{eq:56}
  \bigl\langle (y - \bar y)^k \bigr\rangle &= \int_0^\infty p_y(y)
  \bigl( y - \bar y \bigr)^k \, \diff y \; .
\end{align}
In other words, if we are able to evaluate the moments of $y$ we can
obtain an expansion of $C(w)$.  Actually, the ``centered'' moments can
be calculated from the ``un-centered'' ones, defined by
\begin{equation}
  \label{eq:57}
  \langle y^k \rangle = \int_0^\infty p_y(y) y^k \, \diff y = (-1)^k
  Y^{(k)}(0) \; .
\end{equation}
Here we have used the notation $Y^{(k)}(0)$ for the $k$-th derivative
of $Y(s)$ evaluated at $s = 0$.  Using Eq.~\eqref{eq:19} we can
explicitly write the first few derivatives
\begin{align}
  \label{eq:58}
  Y(0) = {} & 1 \; , \\
  \label{eq:59}
  Y'(0) = {} & \rho Q'(0) \, , \\
  \label{eq:60}
  Y''(0) = {} & \rho Q''(0) + \rho^2 \bigl[ Q'(0) \bigr]^2 \; , \\
  \label{eq:61}
  Y'''(0) = {} & \rho Q'''(0) + 3 \rho^2 Q''(0) Q'(0) + \rho^3 \bigl[
  Q'(0) \bigr]^3 \, , \\
  \label{eq:62}
  Y^{(4)}(0) = {} & \rho Q^{(4)}(0) + 4 \rho^2 Q'''(0) Q'(0) + 3 \rho^2
  \bigl[ Q''(0) \bigr]^2 \nonumber\\
  & {} + 6 \rho^3 Q''(0) \bigl[ Q'(0) \bigr]^2 + \rho^4 \bigl[ Q'(0)
  \bigr]^4 \; .
\end{align}
A nice point here is that, in principle, we can evaluate all the
derivatives of $Y(s)$ in terms of derivatives of $Q(s)$ without any
technical problem.  Moreover, the derivatives of $Q(s)$ in zero are
actually directly related to the moments of $w$.  In fact we have
\begin{equation}
  \label{eq:63}
  Q^{(k)}(0) = (-1)^k \int_\Omega \bigl[w(\vec\theta) \bigr]^k \, \diff^2
  \theta = (-1)^k S_k \; .
\end{equation}
This simple relation allows us to express the moments of $y$ in terms
of the moments of $w$.  For the first ``centered'' moments we find in
particular
\begin{align}
  \label{eq:64}
  \bar y = \langle y \rangle &= \rho \; , &
  \bigl\langle (y - \bar y)^2 \bigr\rangle &= \rho S_2 \; , \\
  \label{eq:65}
  \bigl\langle (y - \bar y)^3 \bigr\rangle &= \rho S_3 \; , &
  \bigl\langle (y - \bar y)^4 \bigr\rangle &= \rho S_4 + 3 \rho^2
  S_2^2 \; .
\end{align}
Hence, we finally have
\begin{align}
  \label{eq:66}
  C(w) \simeq {} & \frac{\rho}{\rho + w} + \frac{\rho^2 S_2}{(\rho
    + w)^3} - \frac{\rho^2 S_3}{(\rho + w)^4} \nonumber\\
  & {} + \frac{\rho^2 S_4 + 3 \rho^3 S_2^2}{(\rho + w)^5} \; .
\end{align}
The first term if this expansion, $\rho / (\rho + w)$, has already
been obtained in Eq.~\eqref{eq:51}.  Other terms represents higher
order corrections to $C(w)$.  In Fig.~\ref{fig:2} we show the result
of applying this expansion to a Gaussian weight.

In closing this section we note that, regardless to the value of
$\pi_w$, $P_0 = \e^{-\rho \pi_w}$ is vanishing at all orders for $\rho
\rightarrow \infty$, and thus we cannot see the peculiarities of
finite-support weight functions here.

\section{Examples}
\label{sec:examples}

In this section we consider three typical examples of weight
functions, namely a top-hat function, a Gaussian, and a parabolic
weight function.  For simplicity, in the following we will consider
weight functions with fixed ``width.''\@  The results obtained can
then be adapted to weight functions with different widths using the
scaling property (Sect.~\ref{sec:scaling}).

\subsection{Top-hat}
\label{sec:top-hat}

The simplest case we can consider for $w(\vec\theta)$ is a top-hat
function of unit radius, which can be written as
\begin{equation}
  \label{eq:67}
  w(\vec\theta) = \frac{1}{\pi} \mathrm{H}\bigl( 1 - | \vec\theta |
  \bigr) \; . 
\end{equation}
In this case we immediately find for $w > 0$
\begin{align}
  \label{eq:68}
  Q(s) & {} = \pi \bigl( \e^{-s/\pi} - 1 \bigr) \; , \\
  \label{eq:69}
  Y(s) & {} = \exp \bigl[ \pi\rho \bigl( \e^{-s/\pi} - 1 \bigr) \bigr]
  \; .
\end{align}
We now note that since $w(\vec\theta)$ is either $0$ or $1/\pi$, we
just need to evaluate $C(1/\pi)$.  We then find
\begin{equation}
  \label{eq:70}
  C(1/\pi) = \frac{\rho}{1 - P_0} \int_0^\infty \e^{-s/\pi} Y(s) \,
  \diff s = 1 \; ,
\end{equation}
as expected.

For the top-hat function we can also explicitly obtain the probability
distribution for $y$.  If $w > 0$ we have
\begin{equation}
  \label{eq:71}
  p_y(y) = \sum_{n = 0}^\infty \frac{\e^{-\rho \pi} (\rho \pi)^n}{\pi
  n!} \delta(y - n/\pi) \; ,
\end{equation}
and thus
\begin{align}
  \label{eq:72}
  C(w) & {} = \frac{\rho}{1 - P_0} \int_0^\infty \frac{p_y(y)}{w + y}
  \, \diff y \notag\\
  & {} = \frac{\rho P_0}{\pi (1 - P_0)} \sum_{n=0}^\infty \frac{(\rho
    \pi)^n}{(w + n/\pi) n!} \; .
\end{align}
From this expression we easily obtain $C(1/\pi) = 1$.  Moreover, we can
evaluate the two limits
\begin{align}
  \label{eq:73}
  \lim_{w \rightarrow \infty} w C(w) & {} = \frac{\rho}{1 - P_0} \; ,
  \\
  \label{eq:74}
  \lim_{w \rightarrow 0^+} w C(w) & {} = \frac{\rho P_0}{1 - P_0} \; ,
\end{align}
thus regaining the results of Sect.~\ref{sec:behavior-w-cw}. 

\subsection{Gaussian}
\label{sec:gaussian}

\begin{figure}[!t]
  \begin{center}
    \resizebox{\hsize}{!}{\includegraphics{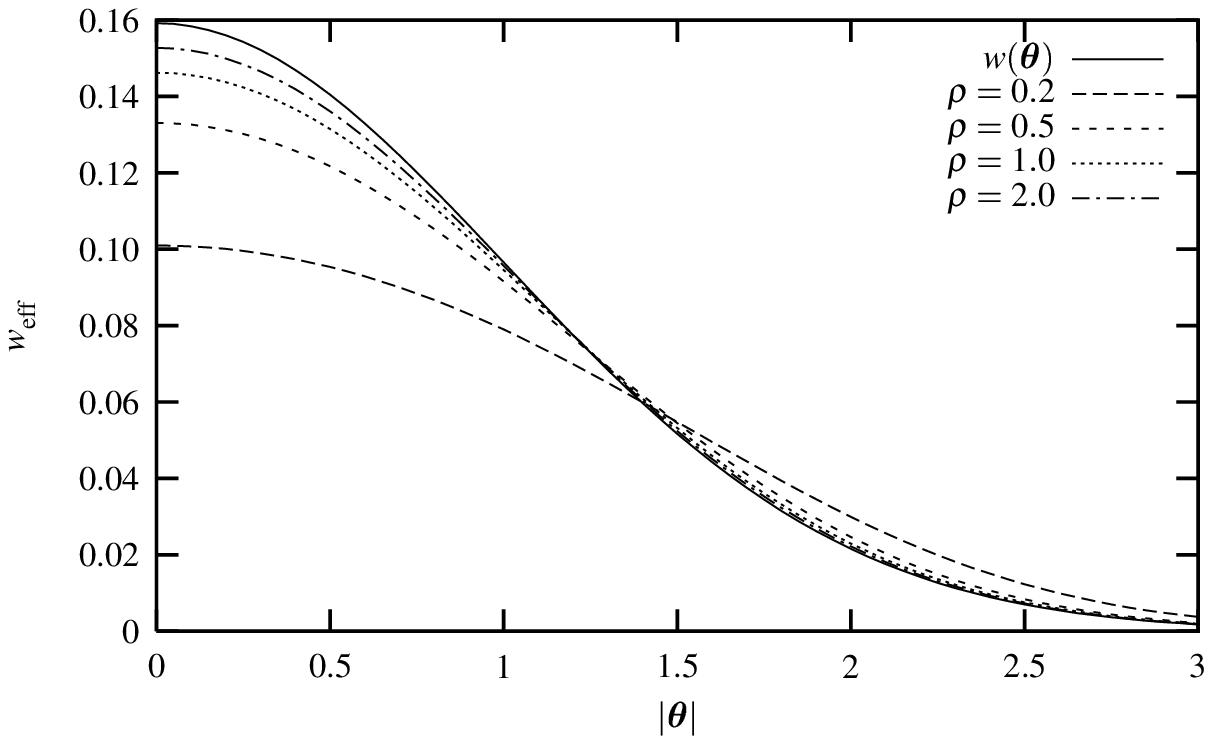}}
    \caption{Effective weight function corresponding to a Gaussian
      weight function.  The original weight function is a normalized
      Gaussian of unit variance, with weight area $\mathcal{A} \simeq
      12.5$.  Significantly broader effective weights are obtained if
      the weight number is smaller than $\mathcal{N} < 10$. }
    \label{fig:3}
  \end{center}
\end{figure}

A weight function commonly used is a Gaussian of the form
\begin{equation}
  \label{eq:75}
  w(\vec\theta) = \frac{1}{2 \pi} \exp\bigl( - | \vec\theta |^2 / 2
  \bigr) \; .
\end{equation}
Unfortunately, we cannot explicitly integrate $Q(s)$ and thus we are
unable to obtain a finite expression for $C(w)$.  The results of a
numerical calculations are however shown in Fig.~\ref{fig:3}.

\subsection{Parabolic weight}
\label{sec:parabolic-weight}

\begin{figure}[!t]
  \begin{center}
    \resizebox{\hsize}{!}{\includegraphics{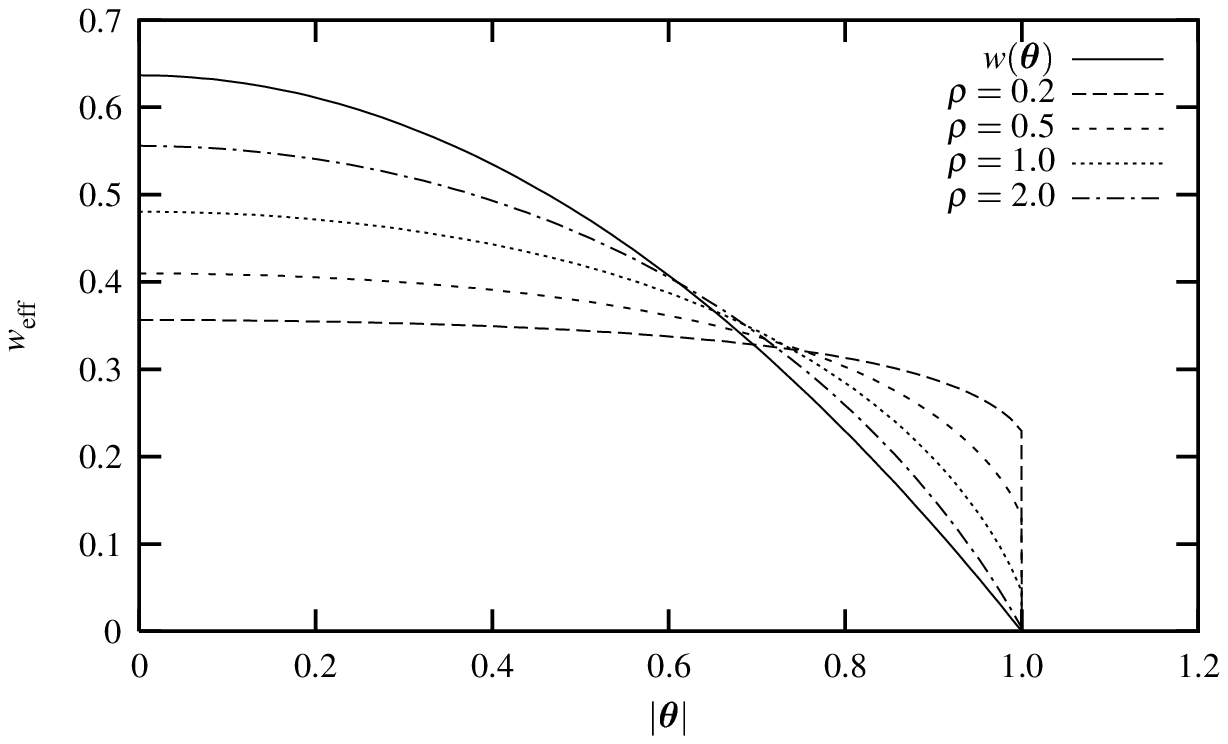}}
    \caption{Effective weight function corresponding to a Gaussian
      weight function.  Note the discontinuity at $| \vec\theta | =
      1$, corresponding to the boundary of the support of $w$.  The
      original weight function is a normalized parabolic function with
      weight area $\mathcal{A} \simeq 2.4$.  Note that this weight
      area is significantly smaller than the ones encountered in
      previous examples.  For low densities, $w_\mathrm{eff}$
      converges to a top-hat function, in accordance with the results
      of Sect.~\ref{sec:limit-large-dens}.}
    \label{fig:4}
  \end{center}
\end{figure}

As the last, example we consider a parabolic weight function with
expression
\begin{equation}
  \label{eq:76}
  w(\vec\theta) = 
  \begin{cases}
    \dfrac{2}{\pi} \bigl( 1 - | \vec\theta |^2 \bigr) & \text{for $|
    \vec\theta | < 1 \; ,$} \\
    0 & \text{otherwise$\; .$}
  \end{cases}
\end{equation}
We then find
\begin{equation}
  \label{eq:77}
  Q(s) = \frac{1 - \e^{-2s/\pi}}{2 s} - \pi \; .
\end{equation}
Unfortunately, we cannot proceed analytically and determine $C(w)$.
We thus report the results of numerical integrations in
Fig.~\ref{fig:4}.  Note that, as expected, the resulting effective
weight has a discontinuity at $| \vec\theta | = 1$.

\section{Conclusions}
\label{sec:conclusions}

In this paper we have studied from a statistical point of view the
effect of smoothing irregularly sampled data.  The main results
obtained can be summarized in the following points.
\begin{enumerate}
\item The mean smooth map, $\bigl\langle \tilde f(\vec\theta)
  \bigr\rangle$, is a \textit{convolution\/} of the unknown field
  $f(\vec\theta)$ with an \textit{effective weight\/}
  $w_\mathrm{eff}$.
\item We have provided simple expressions to evaluate the effective
  weight.  These expressions can be easily used, for example, to
  obtain numerical estimates of $w_\mathrm{eff}$.
\item The effective weight $w_\mathrm{eff}(\vec\theta)$ and the weight
  function $w(\vec\theta)$ share the same support and have a similar
  ``shape.''\@ However, $w_\mathrm{eff}$ is broader than $w$,
  expecially for low densities of objects; moreover, $w_\mathrm{eff}$
  has a discontinuity on the boundary of the support.
\item We have shown that the density of objects $\rho$ (or
  $\rho/(1-P_0)$ for finite-support weight functions) is a natural
  upper limit for the effective weights.
\item The weight number $\mathcal{N}$ has been shown to be the key
  factor that controls the convergence of $w_\mathrm{eff}$ to $w$.  We
  have also shown that $\mathcal{N}_\mathrm{eff} > 1$.
\item The effective weight converges to $w(\vec\theta)$ for large
  densities $\rho$, and to a top-hat function for low densities.
\item We have provided an analytic expansion for $w_\mathrm{eff}$
  which is shown to converge quickly to the exact weight function.
\item Finally, we have considered three typical examples and shown the
  behavior of $w_\mathrm{eff}$ for different densities.
\end{enumerate}
Given the wide use in astronomy of the smoothing technique considered
in this paper, an exact statistical characterization of the
expectation value of the smoothed map is probably interesting
\textit{per se}.  

Finally, we notice that other methods different from Eqs.~\eqref{eq:5}
or \eqref{eq:3} can be used to obtain continuous maps from irregularly
sampled data.  In particular, triangulation techniques can represents
an interesting alternative to the simple weighted average considered
here (see, e.g., Bernardeau \& van de Weygaert 1996; Schaap \& van de
Weygaert 2000).

\begin{acknowledgements}
  We would like to thank the Referee for comments and suggestions that
  enabled us to improve this paper.
\end{acknowledgements}

\appendix

\section{Object intrinsic weight}
\label{sec:object-intr-weight}

The simple weighting scheme considered in this paper [see
Eq.~\eqref{eq:5}] is actually the base of several similar schemes with
slightly different properties.  In this paper, for simplicity and
clarity, we have confined the discussion to the simple case, which is
already rich of peculiarities and unexpected results (for example, the
behavior in case of vanishing weights).  In this appendix, however, we
will describe a slightly more complicated estimator often used in
astronomy.

Suppose, as for the Eq.~\eqref{eq:5}, that we can measure a given
field $f(\vec\theta)$ at some positions $\vec\theta_n$ inside a set
$\Omega$.  Suppose also that we decide to use, for each object, a
weight $u_n > 0$ which is independent of the position $\vec\theta_n$
(the weight $u_n$ could however depend on other intrinsic properties
of the object, such as its luminosity or angular size).  We then
replace Eq.~\eqref{eq:5} with
\begin{equation}
  \label{eq:78}
  \tilde f(\vec\theta) = \dfrac{\sum_{n=1}^N \hat f_n
  u_n w(\vec\theta - \vec\theta_n)}{\sum_{n=1}^N u_n w(\vec\theta -
  \vec\theta_n)} \; .
\end{equation}
A typical situation for which this estimator is useful, is when some
error estimates $\{ \sigma_n \}$ are available for each
object.  Then, if we set $u_n = 1/\sigma_n^2$, we obtain an estimator
which optimizes the signal-to-noise ratio.

Since the quantities $\{ u_n \}$ do not depend on the position, we can
study the statistical properties of the estimator \eqref{eq:78}
provided that the probability distribution $p_u(u)$ of $u$ is
available.  In particular we obtain [cf.\ Eq.~\eqref{eq:8}]
\begin{align}
  \label{eq:79}
  \bigl\langle \tilde f(\vec\theta) \bigr\rangle & {} = \frac{N}{A^N}
  \int_\Omega \diff^2 \theta_1 \int_0^\infty \diff u_1 \, p_u(u_1)
  \, \dotsi \nonumber\\
  & {} \times \int_\Omega \diff^2 \theta_N \int_0^\infty \diff u_N \,
  p_u(u_N) \dfrac{f( \vec\theta_1 ) u_1 w( \vec\theta -
    \vec\theta_1)}{\sum_{n=1}^N u_n w( \vec\theta - \vec\theta_n )} \;
  .
\end{align}
Similarly to Eq.~\eqref{eq:9}, we define
\begin{equation}
  \label{eq:80}
  y(\vec\theta) = \sum_{n=2}^N u_n w(\vec\theta - \vec\theta_n) =
  \sum_{n=2}^N v_n \; ,
\end{equation}
with $v_n = u_n w(\vec\theta_n)$.  Using $y$ and $\bigl\{ v_n \bigr\}$
and defining the probability distributions $p_y(y)$ and $p_v(v)$, we
can write the analogous of Eqs.~\eqref{eq:10} and \eqref{eq:11}.  In
this way we obtain results similar to Eqs.~(\ref{eq:12}-\ref{eq:15})
that we do not report here.  Finally then we have
\begin{align}
  \label{eq:81}
  V(s) & {} = \Lp[p_v](s) = \int_0^\infty \e^{-sv} p_v(v) \, \diff v
  \nonumber\\
  & {} = \frac{1}{A} \int_0^\infty p_u(u) \, \diff u \int_\Omega
  \e^{-s u w(\vec\theta)} \, \diff^2 \theta \; , \\
  \label{eq:82}
  Y(s) & {} = \Lp[p_y](s) = \int_0^\infty \e^{-sy} p_y(y) \, \diff y
  = \bigl[ V(s) \bigr]^{N-1} \; .
\end{align}
In other words, we basically recover Eqs.~\eqref{eq:16} and
\eqref{eq:17} with the significant difference that now the Laplace
transform $V(s)$ of $p_v(v)$ is given by the superpositions of
functions like $W(us)$ weighted with the probability distribution
$p_u(u)$.  Note that all functions $W(us)$ have the same shape but
differ for the scaling of the independent variable.  In case where
$p_u(u) = \delta(u - 1)$ is Dirac's delta distribution centered in
$1$ we exactly reproduce Eqs.~\eqref{eq:16} and \eqref{eq:17}, as
expected.

The continuous limit (cf.\ Sect.~\ref{sec:continuous-limit}) does not
present particular difficulties.  The first significant change regards
Eq.~\eqref{eq:18}, which now becomes
\begin{align}
  \label{eq:83}
  R(s) & {} = \int_0^\infty p_u(u) \, \diff u \int_\Omega \left[
    \e^{-s u w(\vec\theta)} - 1 \right] \, \diff^2 \theta \notag\\
  & {} = \int_0^\infty p_u(u) Q(us) \, \diff u \; .
\end{align}
Correspondingly, $Y(s) = \exp \bigl[ \rho R(s) \bigr]$.  Finally,
writing the effective weight as $w_\mathrm{eff} = w C(w)$, we have
\begin{align}
  \label{eq:84}
  C(w) & {} = \int_0^\infty p_u(u) B(u w) \, \diff u \; , \\
  B(w) & {} = \Lp[\rho Y](w) \; . 
\end{align}

The equations written here can be used to practically evaluate $B(w)$
(and thus $w_\mathrm{eff}(\vec\theta)$), but also to derive the
properties of the effective weight, as done in for the simple
estimator \eqref{eq:5}.  Here we do not carry out the calculations,
since they are straightforward modifications of the calculations of
Sect.~\ref{sec:properties-cw}; moreover the results obtained are
basically identical to the ones reported in that section.

\end{document}